# Thermodynamic Prediction Enabled by Automatic Dataset Building and Machine Learning

Juejing Liu[1,2,*], Haydn Anderson[3], Noah I. Waxman[3], Vsevolod Kovalev[3], Byron Fisher[1], Elizabeth Li[1], and Xiaofeng Guo[1,2,*]

[1] Department of Chemistry, Washington State University, Pullman, Washington 99164, United States

[2] School of Mechanical and Materials Engineering, Washington State University, Pullman, Washington 99164, United States

[3] School of Electrical Engineering & Computer Science, Washington State University, Pullman, Washington 99164, United States

Email of corresponding authors: juejing.liu@wsu.edu and x.guo@wsu.edu

**Abstract**

New discoveries in chemistry and materials science, with increasingly expanding volume of requisite knowledge and experimental workload, provide unique opportunities for machine learning (ML) to take critical roles in accelerating research efficiency. Here, we demonstrate (1) the use of large language models (LLMs) for automated literature reviews, and (2) the training of an ML model to predict chemical knowledge (thermodynamic parameters). Our LLM-based literature review tool (LMExt) successfully extracted chemical information and beyond into a machine-readable structure, including stability constants for metal cation-ligand interactions, thermodynamic properties, and other broader data types (medical research papers, and financial reports), effectively overcoming the challenges inherent in each domain. Using the autonomous acquisition of thermodynamic data, an ML model was trained using the CatBoost algorithm for accurately predicting thermodynamic parameters (e.g., enthalpy of formation) of minerals. This work highlights the transformative potential of integrated ML approaches to reshape chemistry and materials science research.

**Introduction**

Chemical thermodynamics are fundamental for understanding chemical reactions, proposing novel methods to control these reactions, and predicting chemical equilibria/reactions for new materials. Although scientific breakthroughs occur regularly, contributing to these advances becomes progressively more complex. Typical research project necessitates a comprehensive literature review that should cover the current state of the field and identify knowledge gaps. Subsequently, rigorous experimentation and modeling are performed to fill such gaps or check hypothesis-driven predictions. Both these steps are essential research steps not unique in chemical research, which however, are inherently mentally-intensive and time-consuming.[1, 2] To accelerate scientific discovery and enhance research efficiency, the use of machine learning (ML) are increasingly integrating in the fields of chemistry and materials science.[3-8] ML algorithms can be designed to unveil relationships within diverse datasets, encapsulating these patterns into predictive models, which can then be used to generate new data. Overall, ML-assisted approaches offer powerful and new pathways for accelerating experimental discoveries in chemistry and materials science, and potentially for building high-throughput research capabilities.

One significant application of ML in chemistry and materials science is the prediction of compound stability and chemical thermodynamics. By training models on appropriate datasets, ML can quickly correlate the relationships between chemical or crystallographic structures and their corresponding properties, including thermondynamics,[9, 10] which they are naturally connected but may not be readily used empirically or phenomenically. ML-based predictions typically offer a more computationally affordable alternative to *ab initio* methods (e.g., quantum mechanics / molecular mechanics, QMMM, density functional theory, DFT) or classical methods (e.g., classical molecular dynamics, CMD), for predicting characteristics or functionalities. This approach has been successfully employed to predict properties across various reaction types. For instance, the prediction of complexation reactions and stability constants between metal cations and ligands has garnered considerable interest for ML adoption due to its wide-ranging applications,[11-13] including critical element extraction and separation, environmental remediation, and other industrial processes. Thus, ML has proven effective in predicting thermodynamic properties of materials. The ability to accurately predict the properties of compounds and materials can significantly enhance the efficiency of research in chemistry and materials science. While ML is the powerful tool to further advance chemistry studies, training dataset is the bottleneck, especially when datapoints are from a large amount of literature. Traditionally, some of the chemistry-related datasets, e.g., summarizing synthesizing conditions and properties of chemicals/materials from,[14-16] were built manually by reviewing many publications, which is time consuming and prone to human errors.

The recent emergence of generative AI (GenAI) tools, particularly large language models (LLMs), presents new opportunities for advancing chemical research.[17] A primary application of these models is the autonomous literature review and text data mining (TDM) via retrieval-augmented generation (RAG), leveraging their superior natural language processing capabilities to streamline dataset construction.[14, 18-20] In a typical RAG workflow, an LLM generates a response based on a user's prompt and a provided knowledge base—in this case, embedded text from a document. While LLM-based TDM has been successfully adopted for extracting straightforward information, such as reaction conditions or material properties, significant challenges remain. This work focuses on two such challenges. The first is performing TDM on older papers that were digitized from low-quality scans, where errors introduced during optical character recognition (OCR) cause a discrepancy between the text processed by the LLM and the ground truth. The second challenge arises when documents contain a high density of target information (e.g., numerous chemical formula-property pairs), which can overwhelm the LLM's attention mechanism and hinder accurate extraction. Therefore, this work aims to investigate and present solutions to overcome these specific challenges, thereby expanding the capabilities of LLM-based TDM.

Overall, we hope we provided a perspective to investigate an integrated approach combining LLM-based literature review, structured dataset construction, and ML-based property prediction. The developed TDM tool can leverage LLMs to efficiently extract information from diverse documents and organize them into structured datasets. The efficacy of this tool was evaluated for (1) thermodynamic data, including stability constants for metal cation-ligand interactions and thermodynamic properties of rare-earth element (REE) minerals, and (2) broader data types, including non-STEM data (medical research papers), and non-scientific data (financial reports). Our TDM tool successfully mined relevant information in all areas. Interestingly, extracting thermodynamic data proved more challenging than data from other topics, primarily due to less standardized language, complicated table formats, and writing conventions, particularly in publications pre-dating year 2000. To address this, LLM was further requested to rationalize its decision in extracting certain information. This strategy tunes the attention of models to the correct value and therefore improves the successful rate of extraction. Based on the established thermodynamic dataset, we further trained an ML model using the CatBoost algorithm to predict the formation enthalpy of minerals. This model achieved promising accuracy in predicting enthalpy values, using features such as mineral type, chemical formula characteristics, and constituent elements. Our work demonstrates that integrating various ML tools can accelerate thermodynamic investigations with enhanced efficiency for new chemical discoveries, also with potential in other fields.

**Method**

**Software packages and models in literature review.** Our text data mining pipeline is designed to interact with LLMs by using their API calls. We used Mistral optical character recognition (OCR) model to convert portable document format (PDF) document into markdown format.[21] The interaction with Mistral AI was handled by their dedicated Python package. LiteLLM was used to handle all API interactions with LLMs.[22] Google Gemini 1.5-Pro was used to mine data from the document.[23] Other python packages used in this study include NumPy, Pandas, etc.[24, 25]

**Text data mining code.** We developed Language Model Extractor (LMExt) to extract target information from documents. Due to intellectual property protection, we are not able to show code used to run the extraction pipeline. Please see results and discussion about the workflow of data mining. For using this tool, please visit https://lmextract.com/ .

**Examination of extraction result.** The overall extraction accuracy for stability constant, thermodynamic properties, medical papers, and financial documents is defined as correctly extracted items vs. all available items. In the selected paper comparison, the true positive for the F1 value is that LMExt extracts thermodynamic data for a particular chemical formula. The true negative is that LMExt identifies the chemical formula and knows there is no thermodynamic property associated with the formula. The false positive means the model incorrectly assigns the thermodynamic value to a formula, while the false negative is that the model does not extract the thermodynamic properties for the given chemical formula. The composite scores are from the detailed examination of the extracted information from the selected papers, where the ratio is defined by the number of correctly extracted data vs. all data. We further summarized the composite score in terms of text and numeric features.

**Training ML model to predict enthalpy of formation of mineral.** The study utilized a dataset containing REE mineral compositions and their corresponding thermodynamic properties.[26] The initial dataset included mineral formulas, general and specific mineral types, total atoms per formula unit (pfu), and values of standard enthalpy of formation ($\Delta H°_f$). A comprehensive feature engineering process was implemented to extract physical properties from the chemical compositions by using Mendeleev.[27] In detail, the chemical formulas were parsed using regular expressions to obtain normalized atomic ratios for each element present in the compounds. For standardization purposes, the maximum number of elements per compound was set to seven, with zero-padding applied to compounds containing fewer elements. Element-specific features were extracted from a reference database and mapped to each element in the compounds.

After the feature adjustment and augmentation, a four-step approach was used to create the training dataset. First, the data with missing value(s) were removed. Second, the post-cleaning data was further augmented by one-hot encoding function for categorical features (e.g., general and specific mineral types) and StandardScaler function for numerical features (e.g., pfu and element properties).[28] As the values of standard enthalpy of formation (labels) distribute unevenly, logarithmic transformation was applied to smooth the distribution of data. Lastly, a rotation matrix was applied to enhance feature representation.[29] We developed the model predicting $\Delta H°_f$ of REE compounds by CatBoost algorithm.[30] The model was configured with the hyperparameters shown in **Table S1**, tuned by Optuna.[31] The dataset was split into training (80%) and testing (20%) sets, maintaining the random state for reproducibility.

**Results and Discussion**

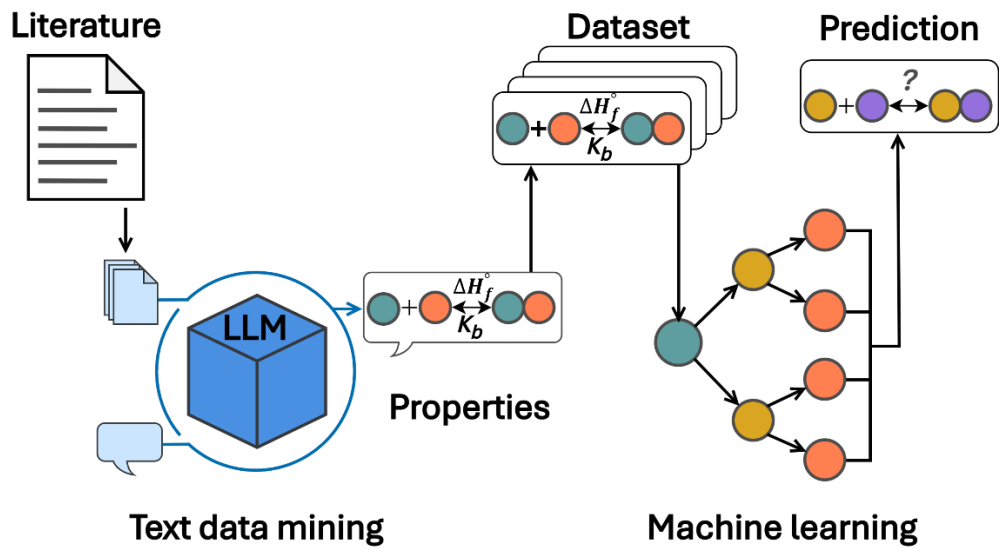

**Figure 1**. Illustration of mining information from scientific literature to build thermodynamic dataset and subsequently perform machine learning predictions.

A literature review pipeline, Language Model Extractor (LMExt, shown in **Figure 1**) was built, to automate the data extraction and cleansing from literature. LMExt uses multiple large language models (LLMs) to extract data from literature with machine-readable format ready for dataset building. We demonstrate the applications of LMExt on extracting datasets on two thermodynamic topics: complexation stability constant and enthalpy and entropy of formation. The generated structured datasets were then used to train a CatBoost model to predict standard enthalpy of formation of minerals.[30]

**Data Mining by LMExt**

LMExt relies on a few-shot learning approach to address the aforementioned challenges. First, Mistral Optical Character Recognition (OCR) model is employed for converting PDF documents into markdown files to circulate the text encoding issues inherent in the PDF files.[21] Then, Google Gemini models are used for separating tables from text and subsequently mining pertinent information.[32] In combination of the prompt instruction, a small size example is input to instruct LLM to extract data in a machine-readable format.

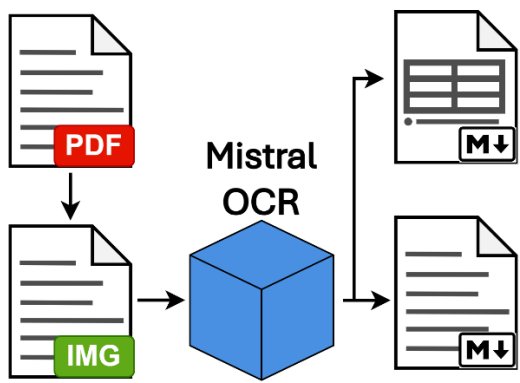

**Figure 2**. Obtaining markdown sibling from literature with PDF format. The PDF documents are firstly converted into images with 400 DPI resolution then loaded into Mistral OCR model. We asked the model to output two markdown documents from one PDF file. One contains text and tables from paper, and another only contains tables in the paper.

For the acquisition of relevant literature, we developed an internal tool that retrieves papers using keyword-based searches. This tool leverages Google Scholar and publisher-provided application programming interfaces (APIs) to perform batch searches and retrieve documents. Although detailed specifications of this internal tool cannot be disclosed, we anticipate that similar functionalities can be readily implemented by referencing the API documentation provided by the publishers.

Conceptually, TDM could be performed by directly uploading PDF documents via an LLM API and using prompts, a method seemingly viable for papers published post-2000.[19] However, for papers published before year of 2000, their PDF documents are highly likely low-quality scans with OCR-based result containing many errors. Moreover, many symbols within PDF documents, including critical ones like $\Delta H°_f$, frequently employ special character encodings. This issue is compounded by the fact that a single symbol can be represented by a wide variety of encoding standards across different publishers, journals, and even different volumes of the same journal. As a result, our preliminary experiments indicated that direct extraction of numerical mineral properties, such as $\Delta H°_f$, from PDF papers, particularly published before 2000, seldom yields satisfactory results. Even advanced multimodal LLMs struggled with direct TDM (i.e., uploading PDF files and querying the LLM), failing to extract information from these legacy documents.

To mitigate the challenges associated with direct PDF processing, we converted the documents into markdown format using the Mistral OCR model (**Figure 2**).[21] This OCR model is specifically tuned to extract text from document images. Treating documents as images circumvents the symbol encoding issues, as the model does not rely on the underlying text encoding to generate a markdown representation. Furthermore, its capability to extract text from low-resolution images is particularly beneficial for processing older papers, which are often low-quality scans. Although the Mistral OCR model API supports direct PDF uploads for markdown conversion, it internally processes PDF pages as images at a 200 dots per inch (DPI, as resolution). While sufficient for papers published after 2000, this resolution proves inadequate for older documents, where the OCR model struggles to accurately recognize subscripts and superscripts (e.g., the '$f$' in $\Delta H_f$, '-1' in mol$^{-1}$, '°' in °C). To address this limitation, we locally converted PDF documents into images at 400 DPI, selected as potentially the highest resolution that avoids triggering Mistral's internal image compression. This step significantly improved OCR accuracy for subscripts and superscripts. Following image conversion, the documents were processed via the Mistral OCR API, from which we requested two markdown outputs: one containing all text and tables, and another containing only the tables. Markdown was chosen as the output format because it effectively balances document simplicity with the preservation of formatting elements like special symbols, subscripts, superscripts, and table structures, thereby markedly improving the success rate of subsequent LLM-based TDM.

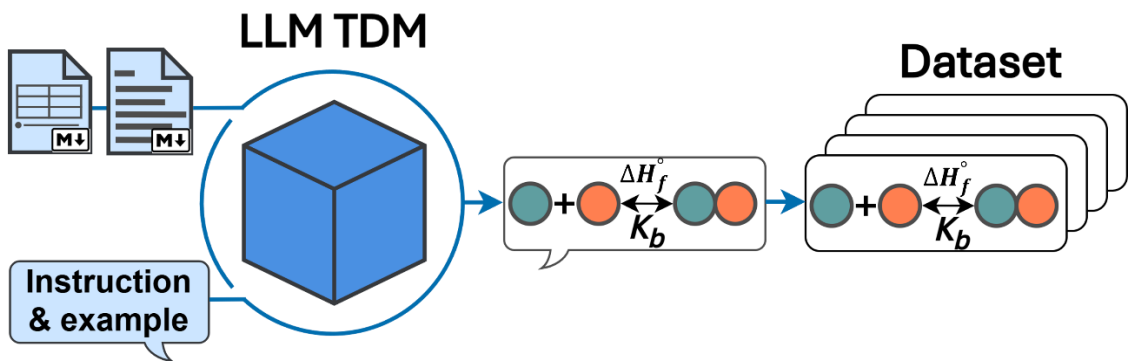

**Figure 3**. Extraction of key parameters from texts and tables in literature by LMExt. The system instruction consists of both command and a few examples to instruct the LLM for mining the correct data and outputting with YAML format. The extracted data is then added to the dataset.

In typical RAG applications, contextual information is commonly pre-embedded, meaning words are converted into corresponding vector representations for the LLM's reference.[33] In contrast, our approach involves directly submitting the content of the markdown files (whether entire papers or isolated tables) to the LLMs (**Figure 3**). The system prompt used for LLM initialization incorporates a few-shot learning approach, instructing the LLM to extract specific information by providing a small number of examples. For instance, when mining enthalpy of formation, the target information includes the chemical formula, phase, and the enthalpy value itself. For complexation constants, the targeted data include cation, anion, pH, temperature, the complexation constant value, among others. These instructions explicitly define the desired output format, while the provided examples illustrate the expected input text and corresponding target output (refer to **Prompt-thermodynamic** and **Prompt-IUPAC** for the specific prompts and examples utilized in Supplementary information). Given that our extraction tasks are well-defined, this direct submission of content elicits responses from the LLM that contain the target information without necessitating an additional prompting step, which would typically be required if pre-embedding were used.

As an initial application, we employed this methodology to construct a digitized, machine-readable dataset of complexation stability constants from a series of PDF documents published by the International Union of Pure and Applied Chemistry (IUPAC).[34] This extensive database contains stability constants ($K_b$) for a wide array of cation-anion interactions. Each PDF document in the series typically covers all cations and anions related to a specific element, with individual anions detailed in separate sections. This particular extraction task was relatively less challenging due to the well-organized, patterned structure of the PDF content. Nevertheless, manual extraction would be exceptionally arduous owing to the sheer volume of data, spanning thousands of pages. The source PDF documents were segmented into numerous smaller sections, each corresponding to distinct anion information, and each section was then processed individually in a separate conversation with the LLM.

**Table 1.** Ratio of successful extraction of information from documents involving complexation stability constant and thermodynamics properties. See result-SC and results-thermodynamic in supporting information for more information. See **stable-constant-comparison** and **thermodynamic-comparison** in Supplementary Information for details.

| Topic | Ratio of successful extraction |
|:---:|:---:|
| Stable constant ($K_b$)* | 100% |
| Thermodynamic ($\Delta H°_f$, $S°_f$), pre-2000 | 43.8% |
| Thermodynamic ($\Delta H°_f$, $S°_f$), post-2000 | 84.2% |

* Examined from 50 pieces of randomly selected extracted vs. ground truth pairs.

Validation against 50 randomly selected data points confirmed that the LLM extracted information from the stability constant database with virtually 100% accuracy (**Table 1**, also see **IUPAC_Results_Raw** in Supplementary Information). Remarkably, the entire corpus, comprising thousands of pages, was processed by a single individual within just a few days. Such levels of accuracy and efficiency are considered unattainable through traditional manual data mining efforts. Thus, even though this particular dataset was relatively straightforward to process due to its well-defined structure, the LLM-based approach significantly enhances the efficiency of generating machine-readable datasets from documents. This capability is crucial for robust dataset preservation and for enabling subsequent data-driven research and machine learning studies.

Extracting thermodynamic properties of minerals from prior publications faced more difficulties compared to mining data from the stability constant dataset. Here we adopted a strategy of separating tables from the main body of the papers. Information was extracted from each paper in a two-pass process: first from a markdown document containing only the tables, and second from the complete paper. This dual approach was implemented again because tables in scientific literature often contain unique structural characteristics distinct from narrative text. Isolating tables enables the design of tailored instructions and examples that better guide the LLMs in extracting the requisite information. Furthermore, despite stylistic variations, tables typically condense a high volume of useful information in a structured format. From an information theory standpoint, tables offer a superior signal-to-noise ratio (SNR) compared to the full text of papers. This higher SNR is advantageous in directing the LLM's attention towards the target information.[35]

Despite the dual-extraction strategy, the initial success rate for extracting thermodynamic data remained low. This difficulty originates from significant variations in terminology, writing logic, and stylistic conventions between papers published before and after the year 2000. Publications from the earlier period, in particular, demonstrate considerable heterogeneity in writing styles, including the use of different terms for the same concept (e.g., "heat of formation" versus "formation enthalpy", "enthalpy of formation", or other similar but not standardized notations) and inconsistent units (e.g., kCal/mol versus kJ/mol).[36, 37] Compounding these issues, the tables within these older papers often exhibit a wide array of non-standardized formats, mirroring the diversity found in the narrative text. Consequently, a robust reasoning capability is crucial for this TDM task, as the model must generalize from a limited set of examples to accurately interpret such diverse presentations of information. However, advanced reasoning models, sometimes referred to as "thinking" models, typically incur high operational costs and exhibit greater latency, rendering them less suitable for large-scale TDM applications.

**Table 2.** Comparison of extraction capability between default extraction and evidence-based extraction. A study published in 1982 studying formation enthalpy of Ln(III) is used as example.[36]

| Cation | Enthalpy of formation (kJ/mol) | |
|:---:|:---:|:---:|
| | Default prompt | Reasoning prompt |
| $La^{3+}$ | - | -708 |
| $Ce^{3+}$ | - | -698 |
| $Pr^{3+}$ | - | -705 |
| $Nd^{3+}$ | - | -696 |
| $Sm^{3+}$ | - | -691 |
| $Eu^{3+}$ | - | -605 |
| $Gd^{3+}$ | - | -687 |
| $Tb^{3+}$ | - | -690 |
| $Dy^{3+}$ | - | -697 |
| $Ho^{3+}$ | - | -706 |
| $Er^{3+}$ | - | -705 |
| $Tm^{3+}$ | - | -702 |
| $Yb^{3+}$ | - | -674 |
| $Lu^{3+}$ | - | -684 |

Incorporating a prompt for evidential support, effectively a "lite" reasoning feature, with a less computationally intensive model (Google Gemini 1.5-Pro) for TDM can significantly improve extraction accuracy, particularly during table processing (**Table 2**). In initial trials where the model was not prompted for evidence, it failed to extract enthalpy of formation from the sample paper, although it did identify the compound types (REE cations). Conversely, when instructed to provide evidence supporting its extraction decisions within the same output (see **Prompt-thermodynamic** in Supplementary Information), the model successfully extracted the enthalpic values from the example paper. This enhancement may be due to the following reasons. First, requiring evidence may prompt the model to engage in a chain-of-thought-like process,[38] compelling it to generate a justification for its extracted answer, thereby implicitly refining its reasoning. Second, the demand for evidence likely constrains the model's output to data that can be explicitly substantiated, ensuring that the generated results fulfill both the user's query and the need for justification, thus reducing the likelihood of irrelevant information being extracted.[39] Third, prompting for evidence may sharpen the model's attention mechanisms, directing its focus more precisely to the relevant data segments within the source text, leading to a higher success rate in extracting information from these evidential sections.[40]

By combining all the above techniques, we mined the thermodynamic properties of REE minerals from literature, including $\Delta H°_f$ and $S°_f$ (see **Table 1,** also see **thermodynamic_mining** in Supplementary Information), from both the pre-2000 and post-2000 papers. The data mining accuracy, defined by whether LMExt extracted the correct value and unit from the paper, for pre-2000 and post-2000 papers are 43.8% and 84.2%, respectively. This result shows that the modern papers published after 2000 are easier to handle, particularly for the OCR model. Therefore, data mining is less challenging and more successful. For the old papers published before 2000, even after careful design of the extraction process and prompts, the accuracy is low. The extraction results contain missing and incorrect values. The low-quality scans in the old papers make the OCR result inaccurate. With an incorrect ORC result, it is impossible to extract the information from the paper.

**Table 3.** Comparison of using different LLMs to mine data from selected papers.[41-45] See **Table S2** for extraction accuracy for every type of data.

| Model | F1 | Composite-numeric | Composite-text |
| --- | --- | --- | --- |
| Gemini-1.5-Pro | 0.65 | 0.89 | 0.93 |
| Gemini-2.5-Flash | 0.77 | 0.60 | 0.81 |
| GPT-4o | 0.75 | 0.71 | 0.95 |
| Llama-3-70b-instruct | 0.62 | 0.82 | 0.90 |
| Claude-3-Opus | 0.28 | 0.64 | 0.58 |
| Gemini-1.5-Flash | 0.57 | 0.92 | 0.44 |

We further evaluated the extraction accuracy across different LLMs on five selected REE mineral thermodynamic papers (see **Table 3** and **Table S2**).[41-45] To compare the performance of different models in terms of both text (e.g. chemical formula) and numeric data (e.g., enthalpic value), two scores were used, namely F1 and Composite. F1 presents whether LMExt identifies the existence of valuable information, while the composite shows how many mistakes (in text and numbers) were presented in the extracted information (see method section for detail). The selection of the papers covers a wide range of sources of data (from tables and/or text) and different writing styles (publication year from 1968 to 2013) in the field of thermodynamic studies. While Gemini-1.5-Pro presents a lower F1 (0.65) score indicating that it did not extract as much information as Gemini-2.5-Flash and GPT-4o, it is the most balanced model in this test. It shows the highest accuracy in the extracted text (0.89) and numbers (0.93). Gemini-2.5-Flash and GPT-4o both identified more information and showed a higher F1 (0.77 and 0.75, respectively). They are less balanced and make more mistakes in numeric extraction. Llama-3 also extracted less information than Gemini-2.5-Flash and GPT-4o (F1 as 0.62). However, the accuracy of extracting numeric and text information is similar (0.82 and 0.90) suggesting its extraction performance is relatively stable as Gemini 1.5-Pro. Claude-3-Opus and Gemini-1.5-Flash perform worse than the above models showing an overall lower F1 and composite.

To further assess the versatility of LMExt, we extended its application to fields beyond chemistry (e.g., medical and financial, see **medical-prompt-result** and **financial-prompt-result** in supplementary information), the pipeline successfully identified extracted all targeted information with 100% accuracy. (e.g., medical studies in the United States and total income from financial reports). These findings indicate that LMExt is highly adaptable across diverse domains, although more extensive experimentation would be beneficial to comprehensively evaluate its performance characteristics.

**ML-based Thermodynamic Property Prediction**

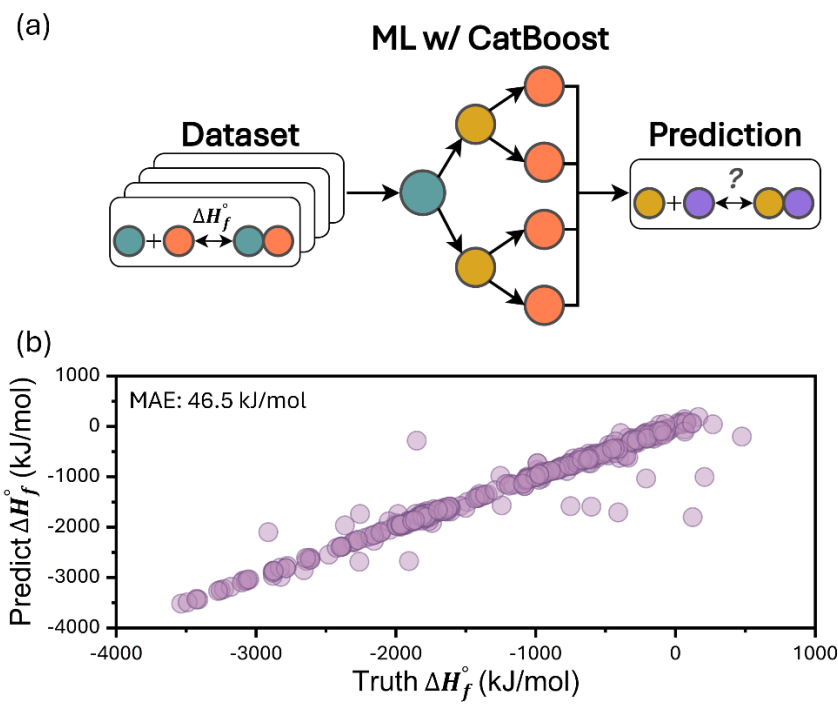

**Figure 4.** Training ML models to predict (a) properties of mineral and demonstration by training a CatBoost model to predict (b) standard enthalpy formation of REE minerals.

Thermodynamic parameters of minerals govern processes such as mineral formation, element speciation and fractionation, and dissolution.[46] From an applied perspective, an understanding of these properties aids in designing separation techniques for extracting valuable elements, including critical elements, from mineral ores.[47] However, the experimental determination of mineral thermodynamic properties is often time-consuming and laborious. Furthermore, some minerals are challenging to synthesize or stabilize under standard laboratory conditions (e.g., phase stability only at high pressures, metastability at ambient conditions, instability when in contact with oxygen or humidity, etc.). Therefore, having predicted thermodynamic properties as a reference prior to experimental measurement is valuable for validating experimental outcomes. In this context, we trained an ML model to predict $\Delta H°_f$ using the CatBoost algorithm (**Figure 4**).[30] The training used our REE thermodynamic dataset, which was constructed through a combination of manual literature review and the LMExt-based results.[26] CatBoost, an algorithm based on the gradient boosting method, is particularly advantageous for handling small datasets. During model training, input features included mineral type, characteristics derived from chemical formulas, and constituent elements, with standard formation enthalpy serving as the target output (refer to the Experimental section for detailed training parameters). The model was configured for regression, with 80% of the 3185 data entries used for training and the remaining 20% allocated for testing model accuracy.

As depicted in **Figure 4b**, the trained model successfully predicted $\Delta H°_f$ of REE minerals within the test dataset, achieving a mean absolute error (MAE) of 46.5 kJ/mol. Notably, 50.5% of these predictions had an absolute error of less than 15 kJ/mol relative to their ground truth values, a degree of error comparable to that typically observed in experimental thermodynamic measurements. These results underscore a significant application of the curated thermodynamic dataset within chemistry and materials science. We anticipate that further enrichment of this dataset will lead to improved prediction accuracy, with the potential for its outputs to eventually serve as a valuable reference for guiding experimental investigations.

**Implication and Perspective**

This study has demonstrated ML-driven solutions and workflows on thermodynamic study in chemistry and materials science targeting literature review and experimental assistance. Our LMExt automates the identification of relevant literature and the summarization of useful information. Crucially, LMExt organizes the extracted data into machine-readable datasets. Such datasets are particularly valuable for subsequent ML applications. In our previous research focused on training ML models to predict outcomes from experimental conditions,[15, 16] the reaction datasets were constructed manually, a process that constituted the most time-consuming aspect of those studies. We anticipate that LMExt will significantly accelerate similar research efforts. Furthermore, the generated data and resulting ML models facilitate data-driven investigations capable of identifying previously unobserved relationships.[16] Although ML models are often perceived as "black boxes," feature importance analysis can still be conducted to understand which input features are most influential for predictions. This capability is particularly significant as it allows for the systematic screening of empirical knowledge, potentially yielding new insights into the subject matter. In summary, we contend that data-driven and ML-assisted chemical research will empower scientists to more effectively uncover fundamental knowledge in thermodynamics of materials.

ML-assisted data analysis has already been widely employed in the fields of chemistry and materials science.[7, 48-54] Depending on the nature of the data requiring analysis, datasets can be generated from experimental and/or synthetic sources to train ML models.[55, 56] When an appropriate dataset is used for training, and the features present in unseen data can be accurately reflected, ML models can efficiently analyze information with minimal human intervention. Furthermore, ML-based data analysis tends to be more objective than manual analysis, provided the training dataset is unbiased. Such rapid and objective data analysis significantly enhances the efficiency of the experimental optimization processes. Moreover, ML-driven data analysis facilitates the integration of high-throughput experimental platforms with reinforcement learning-based experimental optimization.[57, 58] In such configurations, reinforcement learning algorithms can autonomously direct experimental platforms to explore the outcomes of diverse experimental parameters, requiring minimal human oversight until a specific objective is achieved. Highly efficient ML-based data analysis is therefore crucial for the success of these advanced applications.

This research also underscores the significant potential of generative AI in scientific inquiry. Although the current study primarily applied LMExt to review literature within the domains of thermodynamics, the tool itself is designed to be field-agnostic and adaptable to any discipline requiring information extraction from documents. Beyond this specific application, we anticipate that generative AI will assume an increasingly vital role in research by offering novel perspectives derived from the synthesis of extensive datasets.

**Conclusion**

This study demonstrated the synergistic application of large language models and classical machine learning techniques in the field of thermodynamic studies. We developed LMExt, a toolkit that can efficiently extract information from documents and organize it into machine-readable datasets. LMExt was successfully to extract information from different topics: cation-ligand complexation stability constants, thermodynamic properties of REE minerals, medical research papers, and financial reports. The use of prompting LLM has been evidenced in improving data acquisition success rate. Together, we showcase a data mining pipeline that is capable of effectively extracting data from all domains. Furthermore, a machine learning model, trained using the CatBoost algorithm and the curated mineral thermodynamic dataset, accurately predicted the enthalpy of formation of minerals that are not included in its training set. This work illustrates that the integration of ML tools, encompassing both advanced LLMs and classical ML models, can substantially enhance research efficiency. The continued development of this ML-enabled approach holds the potential for high-throughput and autonomous features, relieving researchers from labor-intensive and time-consuming tasks, and allowing us to focus more effectively on the core research to expand the frontiers of scientific knowledge.

**Acknowledgement**

This work was supported by the National Science Foundation (NSF), Division of Earth Sciences, under award No. 2149848, and Division of Materials Research, under award No. 2144792. J.L. and X.G. also acknowledge during the early stage of this work the support by ARPA-E MINER program with award number 0002707-1515. Portions of this research were supported by Alexandra Navrotsky Institute for Experimental Thermodynamics. The authors acknowledge Hannah Hallikainen at WSU for participating in the IUPAC complexation stability constant dataset building.

**Author contributions**

X.G. conceived the project. J.L. and X.G. supervised the project. A.H. developed the data mining code, LMExt. A.H. and V.K. mined data from documents using LMExt. B.F. and V.K. validated data mined from LMExt. E.L. examined dataset for training formation enthalpy prediction model. N.W. trained standard formation enthalpy prediction model. J.L., A.H., N.W., and X.G. drafted and revised the manuscript. All authors have given approval to the final version of the manuscript

**Conflict of interest**

The following authors, X.G. A.H., and J.L., are inventors on intellectual property related to LMExt described in this manuscript, which is owned by Washington State University. The remaining authors declare no competing interests.

**Data availability**

The data presented in this manuscript are available in Supplementary Information and https://github.com/GuoGroupWSU/LLM-thermodynamic. LMExt is accessible through https://lmextract.com/.